# Behavior of Signal Harmonics in Magnetic Particle Imaging Based on a Lock-in-Amplifier Model

Kenya Murase[1,2*]

[1]Department of Medical physics and Engineering, Division of Medical Technology and Science, Course of Health Science, Graduate School of Medicine, Osaka University, Suita, Osaka, Japan
[2]Global Center for Medical Engineering and Informatics, Osaka University, Suita, Osaka, Japan
*Corresponding author, email: murase@sahs.med.osaka-u.ac.jp

**Abstract**
We previously presented a lock-in-amplifier model for analyzing the behavior of signal harmonics in magnetic particle imaging (MPI). In that study, the magnetization and particle size distribution of magnetic nanoparticles (MNPs) were assumed to obey the Langevin theory of paramagnetism and a log-normal distribution, respectively, and a drive magnetic field was assumed to be applied parallel to a selection magnetic field. The purpose of the current study was to investigate the behavior of signal harmonics in MPI with the drive magnetic field being applied to the selection magnetic field in an arbitrary direction using the lock-in-amplifier model. In the lock-in-amplifier model, the signal induced by MNPs in a receiving coil was multiplied with a reference signal, and was then fed through a low-pass filter to extract the DC component of the filtered signal (MPI signal). The strength of the selection magnetic field in MPI was assumed to be given by the product of the gradient strength of the selection magnetic field and the distance from the field-free region (x). The relationships between the MPI signal and x were calculated for odd- and even-numbered harmonics and were investigated for various angles between the axis of a receiving coil and the selection magnetic field. The behavior of signal harmonics in MPI largely depended on the angle between them. This study will be useful for improved understanding, optimization, and development of MPI.

## 1. Introduction

Magnetic particle imaging (MPI) was introduced in 2005 for imaging of the spatial distribution of magnetic nanoparticles (MNPs) [1]. MPI shows great potential in terms of sensitivity, spatial resolution, and imaging speed [1-5]. MPI exploits the nonlinear magnetization response of MNPs to detect their presence in an alternating magnetic field called the drive magnetic field. Spatial encoding is accomplished by saturating the magnetization of the MNPs almost everywhere except in the vicinity of a special region, called the field-free point or field-free line, using a static magnetic field called the selection magnetic field [1-5].

Due to the nonlinear response of the MNPs to an applied drive magnetic field, the signals generated by the MNPs in a receiving coil contain not only the excitation frequency but also the harmonics of this frequency. Because the qualitative and quantitative properties of MPI directly depend on the characteristics of these harmonics, it is important to investigate the behavior of signal harmonics generated by MNPs under various conditions to better understand and optimize MPI.

We previously presented a lock-in-amplifier model for analyzing the behavior of signal harmonics in MPI and reported that the behavior of signal harmonics largely depended on the strength of the drive and selection magnetic fields, the particle size distribution of MNPs, and the parameters in the lock-in amplifier such as the time constant of the low-pass filter [6]. Our previous study [6], however, was limited to the case where the drive magnetic field was applied parallel to the selection magnetic field. The purpose of the current study was to investigate the behavior of signal harmonics in MPI with the drive magnetic field being applied to the selection magnetic field in an arbitrary direction using our lock-in-amplifier model [6], and to present some simulation results. We also present some MPI images obtained using our MPI scanner [7], in which the drive magnetic field is applied perpendicular to the selection magnetic field.

## 2. Materials and Methods

### 2.1. Lock-in-Amplifier Model

As described in our previous paper [6], a lock-in amplifier performs signal mixing, *i.e.*, a multiplication of its input with a reference signal. The mixed signal is then fed through an adjustable low-pass filter to extract the output signal from the DC component of the filtered signal. In this study, the MPI signal ($S_{MPI}$) was defined as the mean absolute value of the output signal [6].

### 2.2. Signals Induced by MNPs

Figure 1 illustrates the relationship between the drive and selection magnetic fields. In Figure 1, $H_S(x)$ is the strength of the selection magnetic field at position $x$, $H_D(t)$ is the strength of the drive magnetic field at time $t$, and $\phi$ is the angle between the axis of a receiving coil and the selection magnetic field.

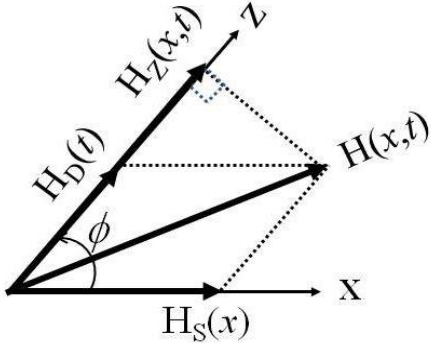

**Figure 1:** Relationship between the drive and selection magnetic fields in simulation studies. $H_D(t)$ is the strength of the drive magnetic field at time $t$ [Equation (6)], $H_S(x)$ is the strength of the selection magnetic field at position $x$ [Equation (7)], and $H(x,t)$ is the strength of the external magnetic field at position $x$ and time $t$ [Equation (5)]. $H_Z(x,t)$ is the z component of $H(x,t)$ [Equation (8)]. $\phi$ is the angle between the axis of a receiving coil and the selection magnetic field.

When MNPs are located at spatial position $r$ and the axis of a receiving coil points in the z direction, parallel to the drive magnetic field, the changing magnetization induces a voltage according to Faraday's law [$v_{rx}(t)$], which is given by [8]

$$v_{rx}(t) = -\mu_0 \frac{\partial}{\partial t} \int_\Omega \sigma_{rx}(\mathbf{r}) \, C(\mathbf{r}) M_Z(\mathbf{r},t) d\mathbf{r} \quad (1)$$

where $\Omega$ denotes the volume containing MNPs, $C(\mathbf{r})$ is the concentration of MNPs at position $\mathbf{r}$, $M_Z(\mathbf{r},t)$ is the z component of the magnetization per unit concentration of MNPs at position $\mathbf{r}$ and time $t$, and $\mu_0$ is the magnetic permeability of a vacuum. $\sigma_{rx}(\mathbf{r})$ is the receiving coil sensitivity derived from the magnetic field that the coil would produce if driven with unit current [8].

In the following, the receiving coil sensitivity is assumed to be constant and uniform over the volume of interest and is denoted by $\sigma_0$. When we consider the signal generated by a point-like distribution of MNPs at position $x_0$, *i.e.*, $C(\mathbf{r}) = C_0 \delta(x - x_0)\delta(y)\delta(z)$ with $C_0$ being constant, the volume integral in Equation (1) vanishes and $v_{rx}(t)$ given by Equation (1) is reduced to

$$v_{rx}(t) = -\mu_0 \sigma_0 C_0 \frac{\partial M_Z(x_0,t)}{\partial t} \quad (2)$$

We neglected constant factors in Equation (2) and $x_0$ was replaced by $x$ for generalization. In addition, we assumed that the signal obtained by the receiving coil includes Gaussian white noise [6].

### 2.3. Langevin Function

Assuming that MNPs are in equilibrium, the magnetization per unit concentration of MNPs at position $x$ and time $t$ [$M(x,t)$] in response to an applied magnetic field can be described by the Langevin function [$\mathcal{L}(\xi)$] [9], which is given by

$$M(x,t) = M_s V_M \mathcal{L}(\xi) = M_s V_M \left( \coth\xi - \frac{1}{\xi} \right) \quad (3)$$

where $M_s$ is the saturation magnetization and $V_M$ is the magnetic volume given by $V_M = \pi D^3/6$ for a particle of diameter $D$. $\xi$ is the ratio of the magnetic energy of a particle with magnetic moment $m$ in an external magnetic field to the thermal energy given by the Boltzmann constant $k_B$ and the absolute temperature $T$, *i.e.*,

$$\xi = \frac{\mu_0 m H(x,t)}{k_B T} = \frac{\mu_0 M_d V_M H(x,t)}{k_B T} \quad (4)$$

In Equation (4), $M_d$ is the domain magnetization of a suspended particle, and $H(x,t)$ is the strength of the external magnetic field at position $x$ and time $t$. As illustrated in Figure 1, $H(x,t)$ is given by

$$H(x,t) = \sqrt{H_S(x)^2 + H_D(t)^2 + 2H_S(x)H_D(t)\cos\phi} \quad (5)$$

In this study, we assumed that $H_D(t)$ is given by

$$H_D(t) = A_D \cos(2\pi f_D t) \quad (6)$$

where $A_D$ and $f_D$ denote the amplitude and frequency of the drive magnetic field, respectively. We also assumed that $H_S(x)$ is given by

$$H_S(x) = G_x \cdot x \quad (7)$$

where $G_x$ and $x$ denote the gradient strength of the selection magnetic field and the distance from the field-free region, respectively.

As illustrated in Figure 1, the z component of $H(x,t)$ is given by

$$H_Z(x,t) = H_D(t) + H_S(x)\cos\phi \quad (8)$$

Thus, we obtain

$$\begin{aligned}\frac{\partial M_z(x,t)}{\partial t} &= M_S V_M \left\{\beta \frac{d\mathcal{L}(\xi)}{d\xi}\left(\frac{H_z}{H}\right)^2 \right.\\ &\left.- \frac{\mathcal{L}(\xi)}{H}\left[\left(\frac{H_z}{H}\right)^2 - 1\right]\right\}\frac{\partial H_z}{\partial t} \\ &= -2\pi f_D A_D M_S V_M \left\{\beta \frac{d\mathcal{L}(\xi)}{d\xi}\left(\frac{H_z}{H}\right)^2 \right.\\ &\left.- \frac{\mathcal{L}(\xi)}{H}\left[\left(\frac{H_z}{H}\right)^2 - 1\right]\right\}\sin(2\pi f_D t)\end{aligned} \quad (9)$$

where $\beta = \mu_0 M_d V_M/(k_B T)$ and $\xi$ is given by Equation (4).

## 2.4. Particle Size Distribution

We assumed that the particle size distribution obeys a log-normal distribution [10]. Thus, the magnetization of MNPs [Equation (3)] averaged based on this particle size distribution ($\langle M(x,t)\rangle$) is given by [11]

$$\langle M(x,t)\rangle = \frac{1}{\sqrt{2\pi}}\int_0^\infty \frac{M(x,t)}{\sigma D}\exp\left[-\frac{1}{2}\left(\frac{\ln(D)-\mu}{\sigma}\right)^2\right]dD \quad (10)$$

In Equation (10), $\mu$ and $\sigma$ denote the mean and standard deviation of the log-normal distribution of diameter $D$, respectively [11].

## 2.5. Simulation Studies

We considered maghemite ($\gamma$-Fe$_2$O$_3$) as MNPs, and $M_d$ in Equation (4) was taken as 414 kA/m [12, 13]. $M_s$ in Equation (9) is given by the product of $M_d$ and the volume fraction of MNPs [12]. The volume fraction of MNPs was assumed to be 1.0 for simplicity. The amplitude and frequency of the drive magnetic field [$A_D$ and $f_D$ in Equation (6), respectively] were fixed at 10 mT and 400 Hz, respectively [2, 3], and the temperature was assumed to be room temperature (293.15 K) in all simulation studies.

The mean diameter of MNPs and $\sigma$ in Equation (10) were assumed to be 20 nm and 0.2, respectively, and $G_x$ in Equation (7) was assumed to be 2 T/m. The time constant of the low-pass filter used in the lock-in amplifier ($\tau$) and the signal-to-noise ratio (SNR) [6] were assumed to be 10 ms and 20, respectively.

## 2.6. Experimental Studies

Experimental studies were performed using our MPI system [7]. In our MPI scanner, a field-free-line selection magnetic field was generated by two opposing neodymium magnets and the gradient strength of the selection magnetic field was 3.9 T/m. An excitation coil for generating the drive magnetic field and a receiving coil were placed between the two neodymium magnets. The amplitude and frequency of the drive magnetic field were 10 mT and 400 Hz, respectively. The drive magnetic field was applied perpendicular to the selection magnetic field. A gradiometer-type receiving coil was used to detect the signal generated by MNPs, which was then transported to a multifunctional filter (3611, NF Co., Yokohama, Japan). The third-harmonic signal was extracted using a lock-in amplifier (LI5640, NF Co., Yokohama, Japan) and converted to digital data using a personal computer with a data acquisition device and a universal serial bus connection (USB-6212, National Instruments Co., TX, USA).

To acquire projection data for image reconstruction, we rotated a phantom placed in the receiving coil through 180° at a sampling angle of 5° and translated it from −16 mm to 16 mm at 1 mm intervals using an XYZ-axes rotary stage (HPS80-50X-M5, Sigma Koki Co., Tokyo, Japan). This stage was controlled using Lab VIEW (National Instruments Co., TX, USA). After projection data were acquired, we transformed each set of projection data into 64 bins by linear interpolation and then reconstructed MPI images using the maximum likelihood-expectation maximization (ML-EM) algorithm with an iteration number of 15 [2, 3, 7].

Resovist® (organ-specific contrast agent for magnetic resonance imaging and $\gamma$-Fe$_2$O$_3$ coated with carboxydextran [12]) (Fujifilm RI Pharma CO., Tokyo, Japan) was used as the source of MNPs.

For phantom experiments, we made a cylindrical acrylic tube phantom (3 mm in inner diameter, 5 mm in outer diameter, 50 μL in volume) filled with Resovist® having an iron concentration of 500 mM (27.9 mg Fe/mL). The $\tau$ value of the low-pass filter used in the lock-in amplifier was varied as 1 ms, 3 ms, 10 ms, 30 ms, 100 ms, and 300 ms.

## 3. Results

Figures 2(a), 2(b), and 2(c) show the relationship between the MPI signal ($S_{MPI}$) and the distance from the field-free region ($x$) for the third, fifth, and seventh harmonics, respectively, for various angles between the axis of a receiving coil and the selection magnetic field ($\phi$), whereas Figures 3(a), 3(b), and 3(c) show those for the second, fourth, and sixth harmonics, respectively. In these cases, $\tau$ and $SNR$ were assumed to be 10 ms and 20, respectively. As shown in Figures 2 and 3, the behavior of signal harmonics largely depended on $\phi$. In the odd-numbered harmonics (Figure 2), the oscillation including the dent characteristic of each harmonic signal decreased with increasing $\phi$. When $\phi = 90°$, no dent was observed. In the even-numbered harmonics (Figure 3), the $S_{MPI}$ value at the peak decreased with increasing $\phi$.

harmonics, and (c) for the seventh harmonics]. The gradient strength of the selection magnetic field ($G_x$), the time constant of the low-pass filter used in the lock-in amplifier ($\tau$), and the signal-to-noise ratio (SNR) were assumed to be 2 T/m, 10 ms, and 20, respectively.

Figure 4(a) shows the MPI images of a cylindrical acrylic tube phantom for various $\tau$ values, whereas Figures 4(b) and 4(c) show the horizontal and vertical profiles through the center of the MPI image for various $\tau$ values, respectively. As shown in Figure 4(a), no artifacts were observed. Furthermore, as shown in Figures 4(b) and 4(c), the profiles showed distributions similar to the Gaussian function in all cases, and blurring was observed when $\tau$ was large.

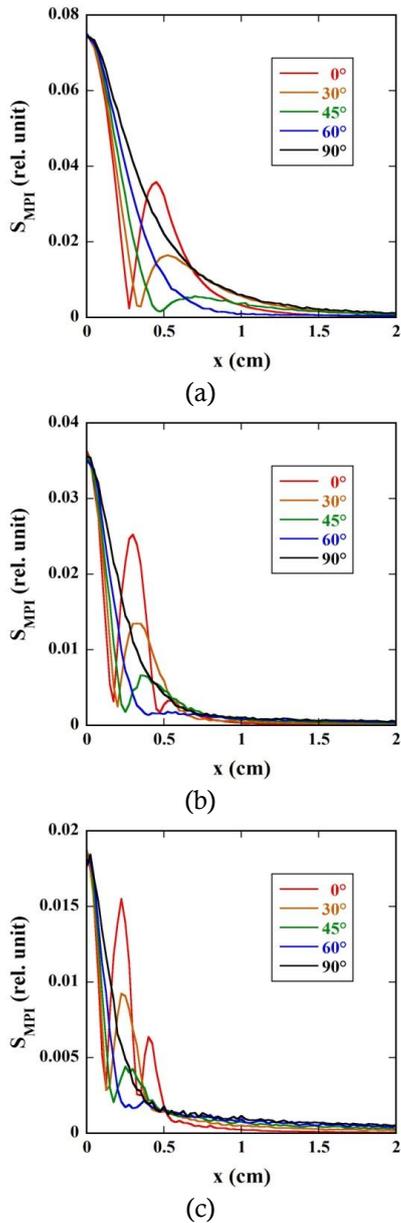

**Figure 2:** Relationship between the MPI signal ($S_{MPI}$) and the distance from the field-free region ($x$) for various $\phi$ values [(a) for the third harmonics, (b) for the fifth

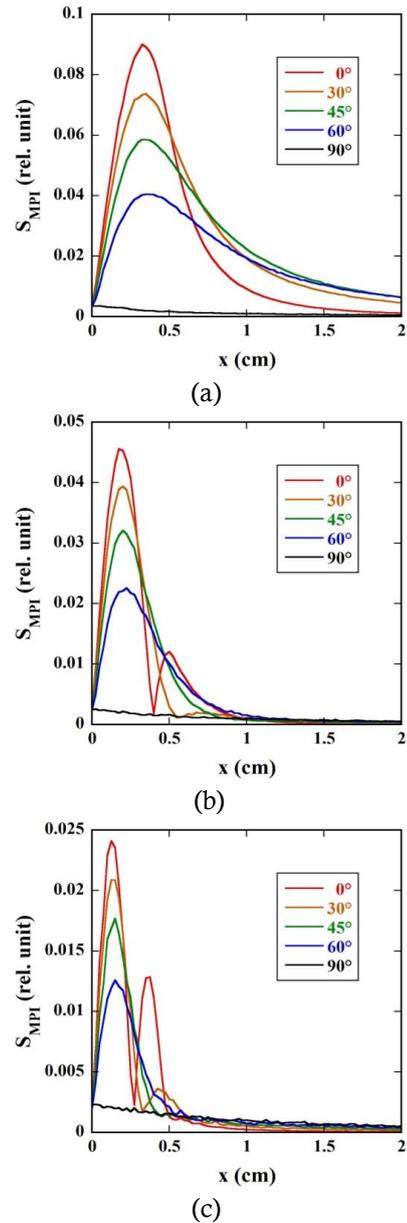

**Figure 3:** Relationship between $S_{MPI}$ and $x$ for various $\phi$ values [(a) for the second harmonics, (b) for the fourth harmonics, and (c) for the sixth harmonics]. $G_x$, $\tau$, and SNR were assumed to be 2 T/m, 10 ms, and 20, respectively.

## 4. Discussion

We previously investigated the behavior of signal harmonics in MPI using the lock-in-amplifier model [6]. In the previous study, the drive magnetic field was assumed to be applied parallel to the selection magnetic field [6]. In this study, we dealt with the case where the drive magnetic field was applied to the selection magnetic field in an arbitrary direction. Our results (Figures 2 and 3) demonstrated that the behavior of signal harmonics in MPI largely depends on the angle between the axis of a receiving coil and the selection magnetic field ($\phi$).

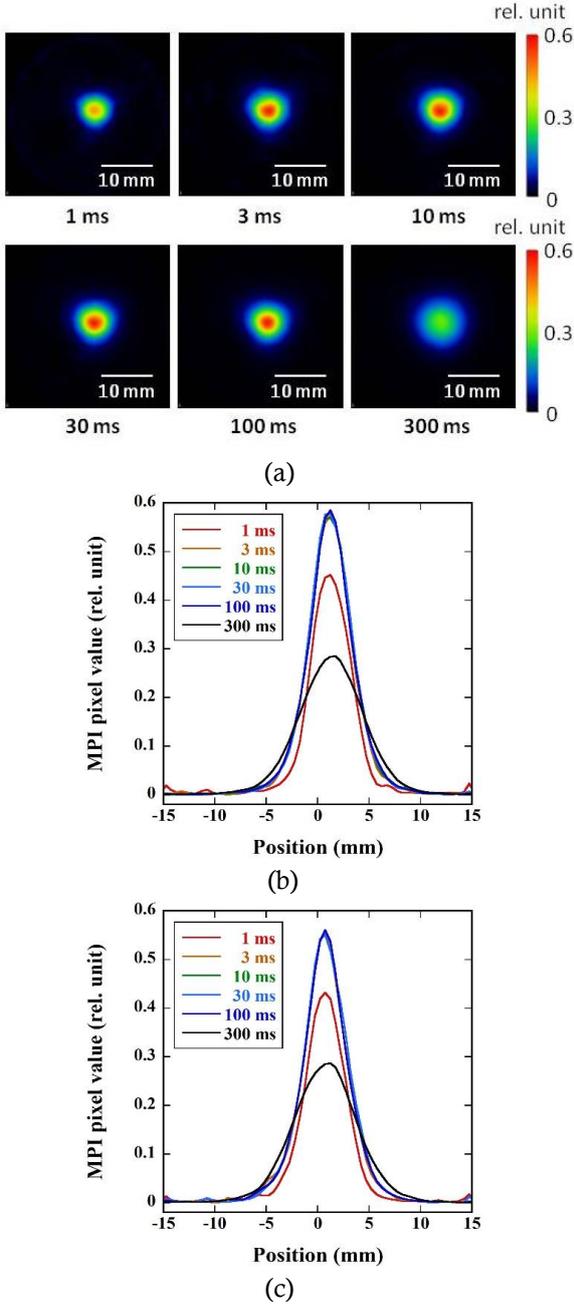

**Figure 4:** (a) MPI images of a cylindrical acrylic tube phantom for various $\tau$ values (1 ms, 3 ms, 10 ms, 30 ms, 100 ms, and 300 ms). Scale bar=10 mm. (b) Horizontal profiles through the center of the MPI image for various $\tau$ values (1 ms, 3 ms, 10 ms, 30 ms, 100 ms, and 300 ms). (c) Vertical profiles through the center of the MPI image for various $\tau$ values (1 ms, 3 ms, 10 ms, 30 ms, 100 ms, and 300 ms).

As shown in Figure 2, when $\phi = 0°$, *i.e.*, the drive magnetic field was applied parallel to the selection magnetic field, oscillation including a dent appeared in the plot of the MPI signal ($S_{MPI}$) versus the distance from the field-free region ($x$) for the odd-numbered harmonics. The oscillation decreased with increasing $\phi$ (Figure 2). When $\phi = 90°$, *i.e.*, the drive magnetic field was applied perpendicular to the selection magnetic field, this oscillation disappeared and $S_{MPI}$ decreased monotonically with increasing $x$. As known from Equation (8), when $\phi = 90°$, $H_Z(x,t)$ becomes equal to $H_D(t)$ and oscillates periodically around $H_Z(x,t) = 0$ irrespective of the selection magnetic field. We believe this explains why the odd-numbered harmonics appear in $S_{MPI}$, while the even-numbered harmonics disappear. Furthermore, $H(x,t)$ given by Equation (5) increases with increasing $x$. Thus, the odd-numbered harmonics appear to decrease monotonically with increasing $x$ as shown in Figure 2. In contrast, when $\phi = 0°$, $H_Z(x,t)$ becomes equal to $H_D(t) + H_S(x)$ and oscillates periodically around $H_Z(x,t) = H_S(x)$. Thus, it appears that the contribution of the odd- and even-numbered harmonics to $S_{MPI}$ changes depending on the selection magnetic field (Figures 2 and 3) and the odd-numbered harmonics do not decrease monotonically with increasing $x$ (Figure 2).

We previously investigated the behavior of signal harmonics in MPI by experimental studies, in which the drive magnetic field was applied parallel to the selection magnetic field [11], and reported that the oscillation including the dent appeared in the plot of $S_{MPI}$ versus $x$ [11]. These results are consistent with those of the present study (Figures 2 and 3). As shown in Figure 4, however, the profiles (corresponding to point spread function) in the MPI images obtained by our MPI scanner decreased monotonically without any dents with increasing distance from the center of the phantom and showed Gaussian-like distributions. This appears to be due to the fact that the drive magnetic field is applied perpendicular to the selection magnetic field in our MPI scanner [2, 3, 7]. These results are also consistent with the simulation results (Figure 2), suggesting the validity of our simulation results. Furthermore, we found that the quality of MPI images depended on the $\tau$ value. These findings are also consistent with our previous simulation results [6].

The relationship between $S_{MPI}$ and $x$ (Figures 2 and 3) corresponds to the system function in the spatial domain in MPI [6]. In the projection-based MPI [2, 3], the projection data are considered to be given by the convolution between the system function in the spatial domain and the line integral of the concentration of MNPs through the field-free line. Thus, the quantitative property of MPI can be enhanced by deconvolution of the system function from the projection data [14]. When we perform such deconvolution, it would be desirable for the system function in the spatial domain to be approximated by a smoothly-changing function such as the Gaussian function. Thus, the present results (Figures 2, 3, and 4) suggest that it is more advantageous to apply the drive magnetic field perpendicular to the selection magnetic field rather than to apply it parallel to the selection magnetic field.

In this study, we analyzed the behavior of signal harmonics in MPI in two dimensions (Figure 1). This analysis can be easily extended to three dimensions, and such studies are in progress.

In conclusion, we presented a method for analyzing the behavior of signal harmonics in MPI with the drive magnetic field being applied to the selection magnetic field in an arbitrary direction and some simulation results based on a lock-in-amplifier model. This method will be useful for improved understanding, optimization, and development of MPI.

**ACKNOWLEDGEMENT**

This work was supported by a Grant-in-Aid for Scientific Research from the Japan Society for the Promotion of Science (JSPS) and the Japan Science and Technology Agency (JST).